\documentclass[reprint,superscriptaddress,nofootinbib,amsmath,amssymb,physrev]{revtex4-2}
\pdfoutput=1
\usepackage{float}
\usepackage{xcolor}
\usepackage{graphicx}
\usepackage{dcolumn}
\usepackage{bm,lipsum}
\usepackage[colorlinks=true, linkcolor=blue, citecolor=blue, urlcolor=blue]{hyperref}
\newcommand{\ZIB}{Zuse Institute Berlin, 14195 Berlin, Germany}
\newcommand{\JCM}{JCMwave GmbH, 14050 Berlin, Germany}

\newcommand{\R}{{\mathbb R}}

\usepackage{graphicx}
\usepackage{booktabs}
\usepackage{tabularx}
\usepackage{xcolor,colortbl}

\definecolor{Gray}{gray}{0.9}

\newcolumntype{a}{>{\columncolor{Gray}}c}
\newcolumntype{b}{>{\columncolor{white}}c}

\begin{document}

\title{Computing eigenfrequency sensitivities near exceptional points}

\author{Felix Binkowski}
\affiliation{\ZIB}
\author{Julius Kullig}
\affiliation{Institut für Physik, Otto-von-Guericke-Universität Magdeburg, 39106 Magdeburg, Germany}
\author{Fridtjof Betz}
\affiliation{\ZIB}
\author{Lin Zschiedrich}
\affiliation{\JCM}
\author{Andrea Walther}
\affiliation{\ZIB}
\affiliation{Department of Mathematics, Humboldt-Universit\"at zu Berlin, 10099 Berlin, Germany}
\author{Jan Wiersig}
\affiliation{Institut für Physik, Otto-von-Guericke-Universität Magdeburg, 39106 Magdeburg, Germany}
\author{Sven~Burger}
\affiliation{\ZIB}
\affiliation{\JCM}

\begin{abstract}
Exceptional points are spectral degeneracies of non-Hermitian systems where both eigenfrequencies and eigenmodes coalesce. The eigenfrequency sensitivities near an exceptional point are significantly enhanced, whereby they diverge directly at the exceptional point. Capturing this enhanced sensitivity is crucial for the investigation and optimization of exceptional-point-based applications, such as optical sensors. We present a numerical framework, based on contour integration and algorithmic differentiation, to accurately and efficiently compute eigenfrequency sensitivities near exceptional points. We demonstrate the framework to an optical microdisk cavity and derive a semi-analytical solution to validate the numerical results. The computed eigenfrequency sensitivities are used to track the exceptional point along an exceptional surface in the parameter space. The presented framework can be applied to any kind of resonance problem, e.g., with arbitrary geometry or with exceptional points of arbitrary order.
\end{abstract}

\maketitle
\section{Introduction}
Resonance phenomena play a crucial role in the field of photonics. The interaction of light with photonic nanoresonators leads to a strong increase of the electromagnetic fields. This effect can be used for, e.g., probing single molecules with ultrahigh sensitivity~\cite{Nie_Science_1997}, designing nanoantennas with a tailored directivity and large spontaneous emission rate~\cite{Askelrod_2014}, or realizing efficient single-photon sources~\cite{Senellart_2017}. Resonances are characterized by loss mechanisms, such as damping or open boundaries, yielding non-Hermitian systems with complex-valued eigenfrequencies~\cite{Lalanne_QNMReview_2018,Zworski_Scattering_Resonances_2019}. In general, the resonances are numerically computed by solving the source-free Maxwell's equations~\cite{Lalanne_QNM_Benchmark_2018,Demesy_ComputPhysComm_2020,Wu_MAN_2023}. For the investigation of photonic systems, not only the resonances are of interest, but also their partial derivatives, the so-called sensitivities, with respect to certain system parameters. These enable a better understanding of the underlying physical effects~\cite{Yan_2020} and an efficient optimization of corresponding photonic devices~\cite{Jensen_LasPhotRev_2011,Binkowski_CommunPhys_2022}.

An impressive signature of the non-Hermitian physics of resonant photonic systems are degeneracies where not only the complex eigenfrequencies but simultaneously also the involved eigenmodes coalesce; for a review see~\cite{Miri_2019}. Commonly, parametric fine-tuning is needed to achieve such non-Hermitian degeneracies, which is why they are also called exceptional points (EPs) in parameter space~\cite{Kato_1995}. In the past years, EPs have been connected to a plethora of interesting effects and suggested applications including ultra-sensitive sensors~\cite{Wiersig_PRL_2014,Wiersig_2016,Chen_Nature_2017,Hodaei_Nature_2017,Lai_Nature_2019,Zeng_OptExp_2019,Hokmabadi_Nature_2019,Wiersig_2020,Park_NatPhys_2020,Kullig_PRR_2021,Kullig_PhotRes_2023}, loss-induced revival of lasing~\cite{Peng_Science_2014}, orbital-angular momentum microlasers~\cite{Miao_Science_2016}, chiral perfect absorbers~\cite{Sweeney_2019}, control of light transport~\cite{Xu_SciAdv_2023}, single-mode lasing~\cite{Hodaei_Science_2014}, electromagnetically induced transparency~\cite{Wang_NatPhys_2020}, optical amplifiers~\cite{Zhong_PRAppl_2020}, and optical filters with sharp transmission peaks~\cite{Tao_OptExp_2010,Chamorro-Posada_2011}. The research on EPs is still a vital field with open problems, such as sensor-performance limitations or Petermann factor divergence at EPs~\cite{Xiao_PRL_2019,Wiersig_NatCom_2020,Wang_NatCom_2020,Kononchuk_Nature_2022,Wiersig_PRR_2023}.

The eigenfrequencies near an EP show a characteristic complex-root topology~\cite{Kato_1995}. Consequently, near the EP, small parametric changes of the system are amplified to a strong response of the eigenfrequencies. This EP-enhanced eigenfrequency sensitivity is challenging in experiments and may also spoil the accuracy of numerical simulations. On the other hand, calculation of the eigenfrequency sensitivities near an EP is crucial for applications such as sensors or optimization schemes. Hence, accurate and efficient numerical methods for photonic systems operating near an EP are of fundamental interest in research and engineering.

\begin{figure}[]
\includegraphics[width=0.49\textwidth]{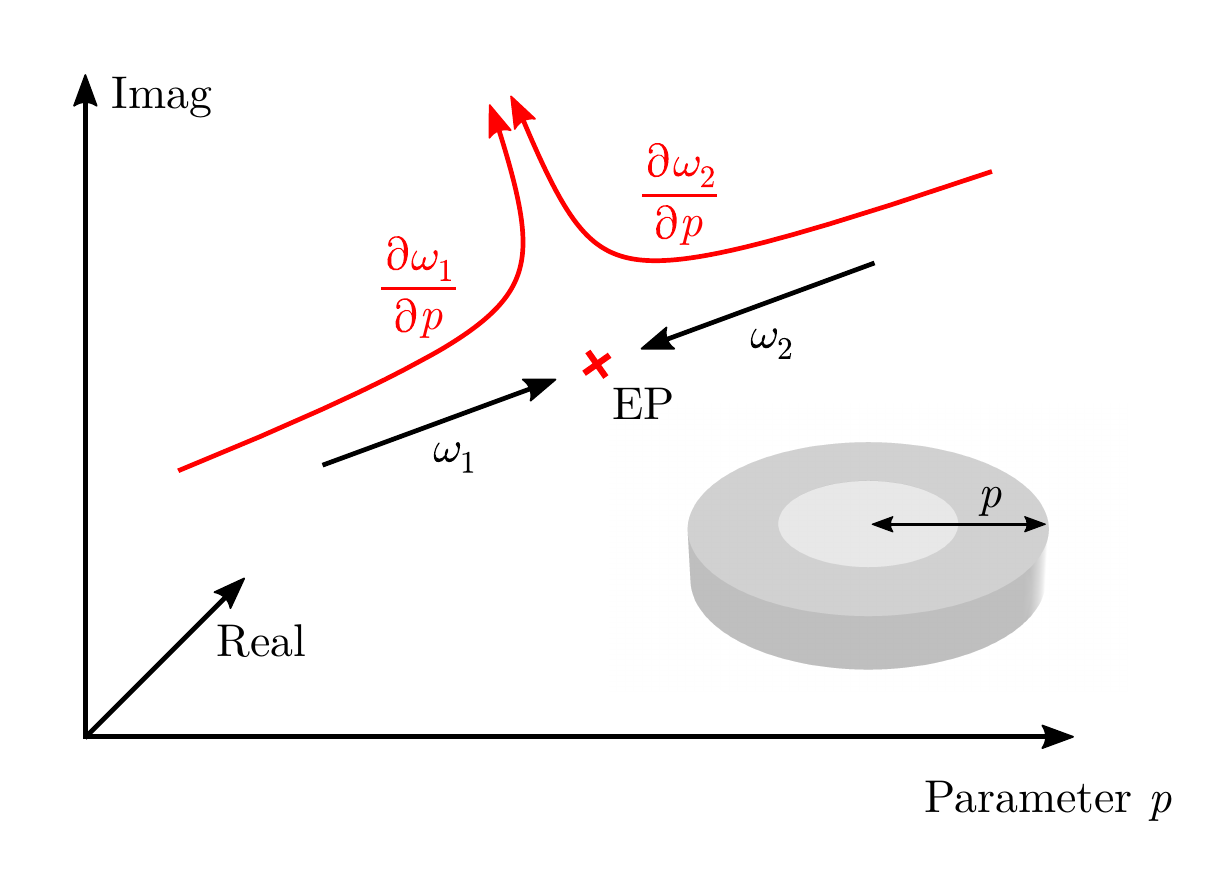}
\caption{\label{fig01}
Eigenfrequencies ${\omega}_l$ and their sensitivities $\partial {\omega}_l / \partial p$ in the complex frequency plane with respect to a parameter $p$ of an optical microdisk cavity. A specific change in the parameter cause the eigenfrequencies and the associated eigenmodes to coalesce at the EP. The sensitivities of the eigenfrequencies diverge near the EP.}
\end{figure}

In this work, we address this challenge by combining contour integration with algorithmic differentiation (AD). Contour integration gives access to eigenfrequencies in non-Hermitian systems by solving Maxwell's equations with source terms at frequencies on a contour in the complex frequency plane.  
For the corresponding scattering problems, AD is exploited so that the calculation of eigenfrequency sensitivities is numerically accurate and efficient. Although the eigenfrequency sensitivities diverge near an EP, which is sketched in Fig.~\ref{fig01}, the presented framework allows to capture the EP-enhanced eigenfrequency sensitivity. We apply the framework to an optical microdisk cavity with an EP of second order. The two coalescing eigenfrequencies and the corresponding sensitivities are computed near the EP. The sensitivities are further used for an optimization scheme to track the EP along a surface in the underlying parameter space. Such exceptional surfaces are used to combine the sensitivity of EPs with robustness against fabrication tolerances \cite{ZRK19,ZNO19,Soleymani_NatCom_2022}.
\vspace{-0.35cm}

\section{Theoretical background} 
In photonics, in the steady-state regime, light scattering by an open system can be described by the time-harmonic Maxwell's equation equipped with open boundary conditions,
\begin{align}
	\nabla \times  \mu_0^{-1} 
	\nabla \times \mathbf{E}(\mathbf{r},\omega)  -
	\omega^2\epsilon(\mathbf{r},\omega) \mathbf{E}(\mathbf{r},\omega)  = 
	i\omega\mathbf{J}(\mathbf{r}), \label{eq:maxwell}
\end{align}
where $\mathbf{E}(\mathbf{r},\omega) \in \mathbb{C}^3$ is the electric field, $\mathbf{J}(\mathbf{r}) \in \mathbb{C}^3$ is a source term corresponding to an optical source, $\omega \in \mathbb{C}$ is the angular frequency, and $\mathbf{r} \in \mathbb{R}^3$ is the spatial position. For optical frequencies, the permeability $\mu(\mathbf{r},\omega) = \mu_\mathrm{r}(\mathbf{r},\omega) \mu_0$ typically equals the vacuum permeability $\mu_0$. The permittivity $\epsilon(\mathbf{r},\omega) = \epsilon_\mathrm{r}(\mathbf{r},\omega) \epsilon_0$, where $\epsilon_\mathrm{r}(\mathbf{r},\omega)$ is the relative permittivity and $\epsilon_0$ is the vacuum permittivity, describes the material dispersion and the spatial distribution of material. Problems given by Eq.~\eqref{eq:maxwell} are called scattering problems. Resonance problems are given by Eq.~\eqref{eq:maxwell} but without a source term. Note that, in the following, we consider non-Hermitian systems based on open boundaries. However, the inclusion of damping as loss channel is also possible.

\subsection{Contour integration for computing eigenfrequency sensitivities near exceptional points} 
Electromagnetic quantities $q(\mathbf{E}(\mathbf{r},\omega)) \in \mathbb{C}$, such as reflection or transmission coefficients, are measured for real excitation frequencies \mbox{$\omega \in \mathbb{R}$}. To compute complex-valued eigenfrequencies and their sensitivities in non-Hermitian systems, we consider the analytical continuation of $q(\mathbf{E}(\mathbf{r},\omega \in \mathbb{R}))$ into the complex frequency plane $\omega \in \mathbb{C}$, which we denote by $q(\omega)$ as a short notation of $q(\mathbf{E}(\mathbf{r},\omega \in \mathbb{C}))$~\cite{Binkowski_PRB_2024}. The $L$ eigenfrequencies inside a chosen contour $C$ are given by the eigenvalues ${\omega}_l$ of the generalized eigenproblem~\cite{Delves_1967,Kravanja_2000,Austin_2014}
\begin{align}
H^{<} X = H X \Omega, \label{eq:Hankel}
\end{align}
where $\Omega = \mathrm{diag}({\omega}_1,\dots,{\omega}_L)$
is a diagonal matrix containing the eigenvalues, the columns of the
matrix $X \in \mathbb{C}^{L\times L}$ are the eigenvectors, and
\begin{align}
	H =
	\begin{bmatrix}
	s_0	& \dots & s_{L-1} 		\\
	\vdots 	&   		 & \vdots	\\
	s_{L-1}		&\dots 	& s_{2L-2}		
	\end{bmatrix}, \,\,\,
  	H^{<} =
	\begin{bmatrix}
	s_1	& \dots & s_{L} 		\\
	\vdots 	&    	 & \vdots	\\
	s_{L}		&\dots 	& s_{2L-1}		
        \end{bmatrix} \nonumber
\end{align}
are Hankel matrices with the contour-integral-based elements
\begin{align}
s_k = \frac{1}{2\pi i} \oint \nolimits_{{C}} \omega^k q(\omega)
     d\omega = \sum_{l=1}^L {\omega}_l^k a_l. \label{eq:hankel_elements}
\end{align}
The second equality results from applying Cauchy's residue theorem, where $a_l$ are the residues of $q(\omega)$ corresponding to the eigenfrequencies ${\omega}_l$, and it holds for simple eigenfrequencies.

Once the elements $s_k$ are computed, there is not only access to the eigenfrequencies by solving the eigenproblem given in Eq.~\eqref{eq:Hankel}, but also to the eigenfrequency sensitivities $\partial {\omega}_l / \partial p$, where $p$ is some parameter. Computing the derivatives in Eq.~\eqref{eq:hankel_elements} yields the linear system of equations for the unknowns $\partial {\omega}_l / \partial p$ and $\partial a_l / \partial p$,
\begin{align}
\begin{split}
\frac{\partial s_k}{\partial p} &= \frac{1}{2\pi i} \oint \nolimits_{{C}} \omega^k \frac{\partial q(\omega)}{\partial p}
     d\omega \\ &=  \sum_{l = 1}^{L}\left(k{\omega}_l^{k-1}\frac{\partial {\omega}_l}{\partial p} a_l + {\omega}_l^k\frac{\partial a_l}{\partial p} \right),  
     \end{split}\label{eq:derivative}
\end{align}
where ${\omega}_l$ and $a_l$ are known due to solving Eq.~\eqref{eq:Hankel} and using $\mathrm{diag}(a_1,\dots,a_L) = X^T H X (V^TX)^{-2}$ with the Vandermonde matrix~\cite{Binkowski_PRB_2024}
\begin{align}
V = \begin{bmatrix} 1 & \dots & 1 \\ {\omega}_1 & \dots & {\omega}_L \\ \vdots & & \vdots \\ {\omega}_1^{L-1} &\dots & {\omega}_L^{L-1} \end{bmatrix}. \nonumber
\end{align}

The eigenfrequencies at EPs are not simple. However, experimental and numerical realizations show always an eigenfrequency splitting due to fabrication inaccuracies~\cite{ZRK19} or numerical approximation~\cite{Wiersig22c}. For this practical reason, we can consider Eq.~\eqref{eq:derivative} to compute eigenfrequency sensitivities in EP-based systems.

\begin{figure}[]
\includegraphics[width=0.49\textwidth]{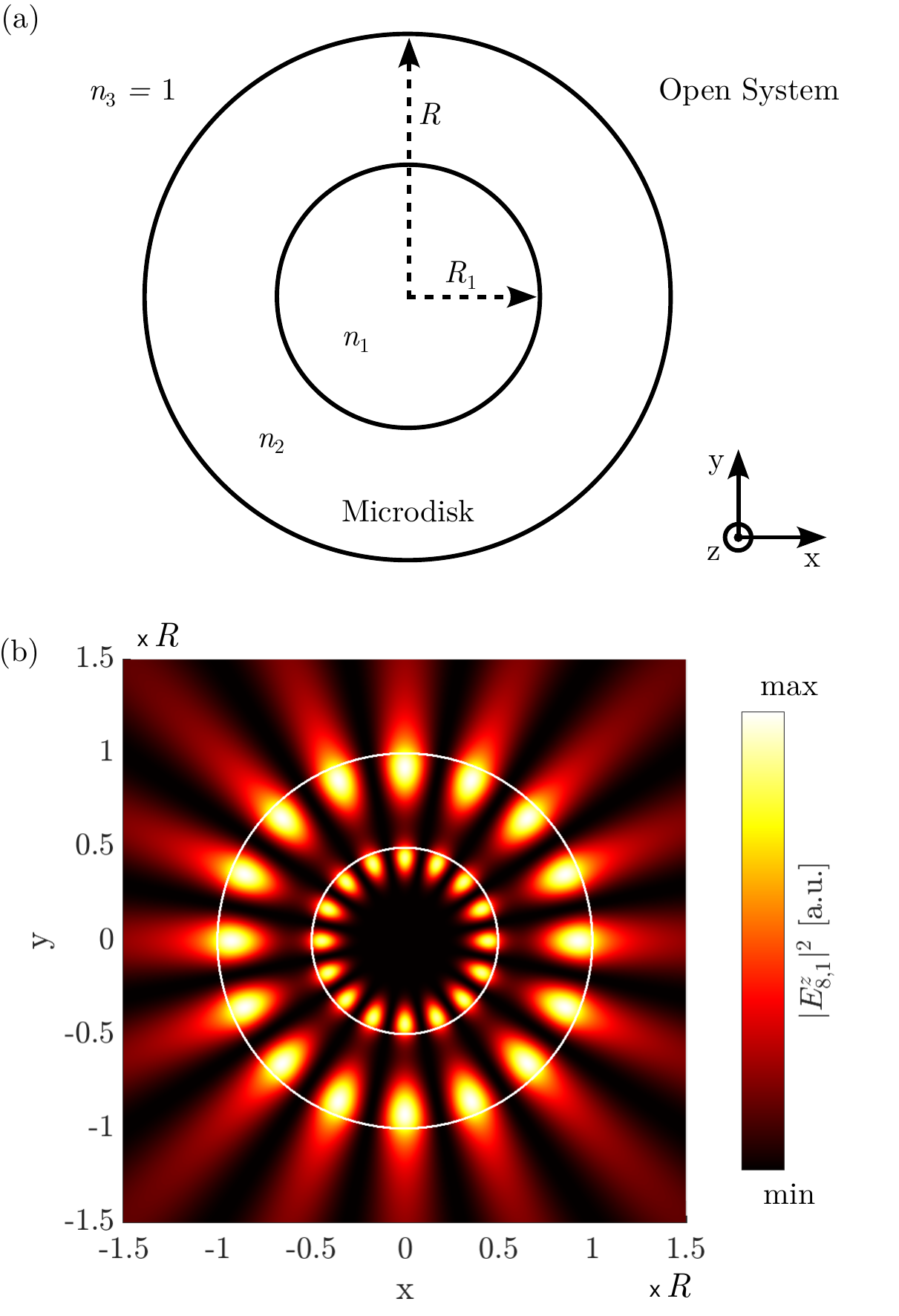}
\caption{\label{fig02}
Optical microdisk cavity with an EP of second order.
(a) Sketch of the two-dimensional system. Two concentric layers of different refractive indices form the cavity.
(b) Electric field intensity of the eigenmode ${E}_{8,1}^z$ corresponding to the EP. The associated eigenfrequency is given by ${\omega}_{8,1} \approx {\omega}_{8,2} = (6.96185 - 0.089761i)\times c/R$. The underlying para\-meters of the microdisk are $n_1 = 3.1239791$, $n_2 = 1.5$, and $R_1 = 0.4970147\times R$.}  \vspace{-0.3cm}
\end{figure}

The elements $s_k$ of the Hankel matrices and their sensitivities $\partial s_k/\partial p$ are obtained by numerical quadrature~\cite{Trefethen_SIAM_Trapz_2014} of the contour integrals in Eq.~\eqref{eq:hankel_elements} and Eq.~\eqref{eq:derivative}, respectively, where $q(\omega)$ and $\partial q(\omega)/ \partial p$ are computed for complex frequencies on the integration contour $C$. The quantities $q(\omega)$ and $\partial q(\omega)/ \partial p$ only have to be computed once for each integration point and then all contour integrals can be evaluated. The integrands differ only in the weight functions $\omega^k$. Information on the numerical realization of the contour integration can be found in the Refs.~\cite{Betz_2021,Binkowski_SourceCode_Sens_EPs}.

\subsection{Structure exploiting algorithmic differentiation}
To evaluate $q(\omega)$ and $\partial q(\omega)/ \partial p$ on the integration contour, the electric field $\mathbf{E}(\mathbf{r},\omega)$ and the corresponding sensitivity $\partial \mathbf{E}(\mathbf{r},\omega)/ \partial p$ have to be computed. For this, we solve Maxwell's equation given in Eq.~\eqref{eq:maxwell} using the finite element method (FEM) with the implementation of the software package \textsc{JCMsuite}~\cite{Pomplum_NanoopticFEM_2007}. The spatial discretization of Eq.~\eqref{eq:maxwell} yields the linear system of equations $AE =f$, where $A \in \mathbb{C}^{n \times n}$ is the FEM system matrix, $E \in \mathbb{C}^{n}$ is the scattered electric field in a finite-dimensional FEM basis, and $f\in \mathbb{C}^{n}$ realizes the source term.

\begin{figure*}[]
\includegraphics[width=0.98\textwidth]{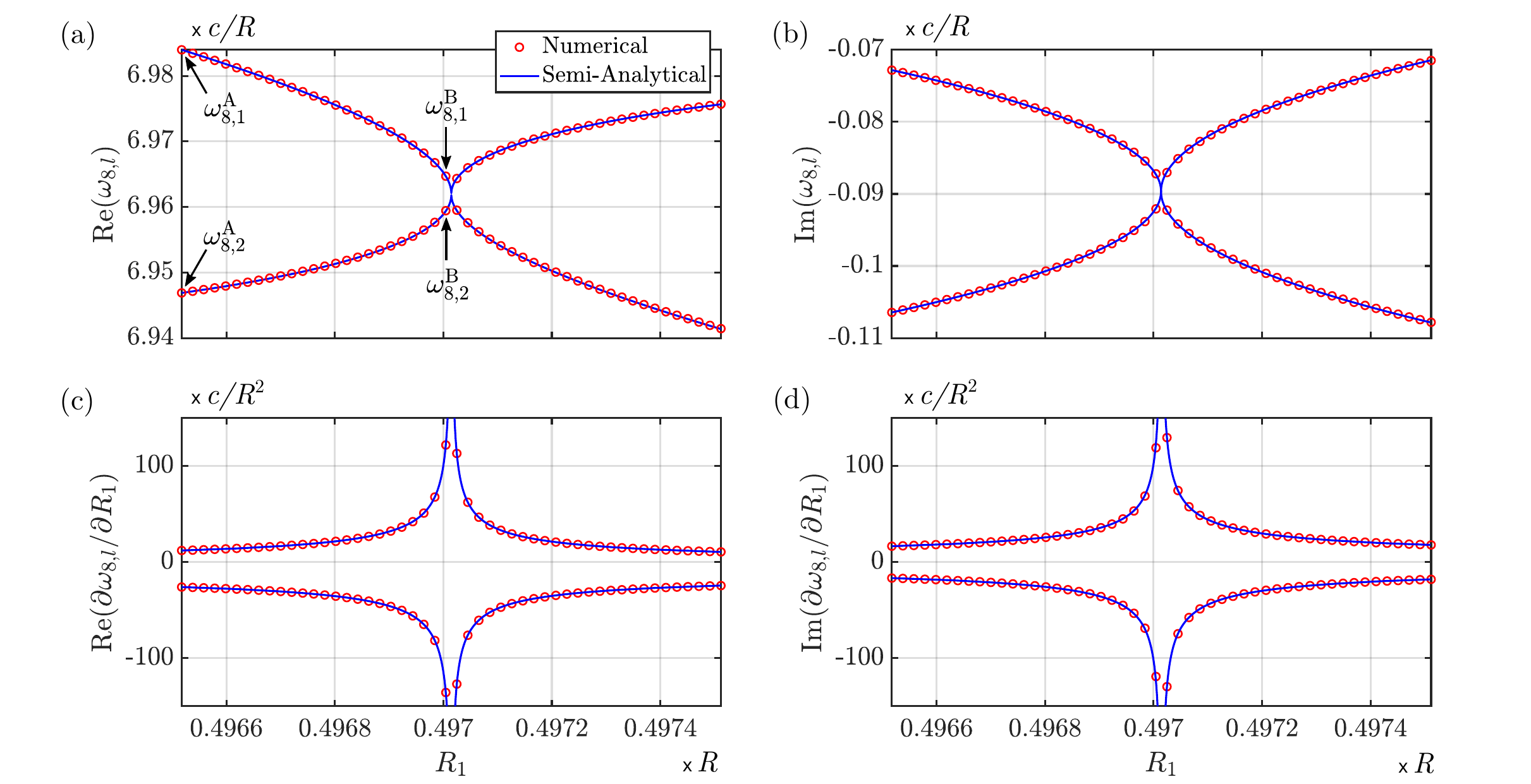}
\caption{\label{fig03}
Variation of the parameter $R_1$ near the EP with
the parameters $n_1 = 3.1239791$, $n_2 = 1.5$, and $R_1 = 0.4970147\times R$.
For each $R_1$, two eigenfrequencies ${\omega}_{8,1}$ and ${\omega}_{8,2}$ are obtained.
The values for $R_1$ result from an equidistant spacing in the interval $[0.999,1.001]\times R_1$ with 50 points, where the first point leads to ${\omega}_{8,1}^\mathrm{A}$ and ${\omega}_{8,2}^\mathrm{A}$ and the $25$th point leads to ${\omega}_{8,1}^\mathrm{B}$ and ${\omega}_{8,2}^\mathrm{B}$. The numerical results can be compared with the semi-analytical solution. 
(a)~Real parts of the eigenfrequencies ${\omega}_{8,l}$.
(b)~Imaginary parts of ${\omega}_{8,l}$.
(c)~Real parts of the eigenfrequency sensitivities $\partial {\omega}_{8,l}/ \partial R_1$.
(d)~Imaginary parts of $\partial {\omega}_{8,l}/ \partial R_1$.
}
\end{figure*}

For the computation of $\partial \mathbf{E}(\mathbf{r},\omega)/ \partial p$, an approach based on directly using the FEM system matrix is applied~\cite{Nikolova_2004,Burger_Derivatives_PROCSPIE_2013,Binkowski_CommunPhys_2022}. With this direct differentiation approach, the sensitivities of scattering solutions can be computed by
\begin{align}
\frac{\partial E}{\partial p} = A^{-1} \left( \frac{\partial f}{\partial p}
- \frac{\partial A}{\partial p}E \right). \nonumber
\end{align}
Instead of directly computing $A^{-1}$, an $LU$-decomposition of $A$ is calculated to efficiently solve the linear system $AE=f$. The $LU$-decomposition is used to obtain $E$ and also ${\partial E}/{\partial p}$. In the FEM context, an $LU$-decomposition is usually a computationally expensive step, so reusing it significantly reduces computational effort~\cite{Binkowski_CommunPhys_2022}.

Several approaches could be used to evaluate the sensitivities of the system matrix, $\partial A/ \partial p$, and of the source term, $\partial f/ \partial p$. The required derivative information can be provided analytically by means of corresponding computer algebra systems if the considered functional dependence is not too complicated. Alternatively, one may employ finite differences to approximate the gradient information. However, the resulting computational effort scales linearly with the number of parameters $p$~\cite{Binkowski_CommunPhys_2022}. Furthermore, finite differences only yield an approximation of the required derivative information, a fact that may cause problems if the accuracy of the derivatives is essential. For this reason, we propose AD to provide the required sensitivities within working accuracy in an efficient way \cite{GrKuWa12}.  Suppose a function ${F}:\R^n \mapsto \R^m$, $y =F(p)$, is given in a computer language like C or C++.  Then, the evaluation of $F(p)$ can be decomposed into so-called elementary functions, the derivatives of which are well known and easy to implement in a software package. The basic differentiation rules, such as the product rule, the quotient rule, etc., can be applied to each statement of the given code segment to calculate the overall derivatives. Hence, exploiting the chain rule yields the derivatives of the whole sequence of statements with respect to the input variables. 

One distinguishes two basic modes of AD, namely the forward mode and the reverse mode. The former one evaluates derivatives together with the function evaluation. In mathematical terms, one obtains, for a given direction $\dot{p}$, the Jacobian-vector product $\dot{y} = {F}'(p) \dot{p}$. Hence, for a unit vector $\dot{p} = e_i\in \R^n$, one obtains the $i$th column of the Jacobian $\nabla {F}(p)$. When applying the reverse mode of AD, one propagates derivative information from the dependents ${y}$ to the independents ${p}$ after the evaluation of the function $F(p)$. This yields, for a given weight vector $\bar{y}$, the vector-Jacobian product $\bar{p} = \bar{y}^\top F'(p)$. Similar to the forward mode, for a unit vector $\bar{y} = e_j\in \R^m $, one obtains the $j$th row of the Jacobian $\nabla {F}(p)$.

Over the past decades, extensive research activities led to a
thorough understanding and analysis of these two basic modes of AD, where
the complexity results with respect to the required runtime are based on the 
operation count $O_{F}$,~i.e.,~the number of floating point
operations required to evaluate ${F} (p)$.
The forward mode of AD yields one {\em column} of the Jacobian $\nabla {F}$
for no more than three times $O_{F}$~\cite{GrWa08,Na12}. Using the reverse mode of AD, 
one {\em row} of $\nabla {F}$ is obtained for no more
than four times $O_{F}$, also see Ref.~\cite{GrWa08}. It is important to note
that this bound for the reverse mode is completely independent of the
number $n$ of input variables. Hence, if $m=1$, the gradient of the then scalar-valued 
function~${F}$ can be calculated for four function evaluations. This observation is called 
{\em cheap gradient result} and is used extensively for
derivative-based optimization approaches. 
However, the reverse mode requires the knowledge of
intermediate results computed during the function
evaluation. Therefore, the basic implementation of the reverse mode
leads to a memory requirement that is proportional to the operation
count $O_{F}$. For a considerable amount of applications, this
fact does not cause any problems. For problems of larger scale,
checkpointing strategies have been developed, see, e.g., Ref.~\cite{revolve}. Here, instead of storing
all intermediates, only a few of them are recorded. Subsequently, the
missing intermediate values are recomputed using the data stored in
the checkpoints. Hence, these checkpointing strategies seek for a 
compromise between storing and recomputing data. 
Besides the theoretical foundation, numerous AD tools have been
developed, e.g.,~CppAD~\cite{cppad}, ADOL-C~\cite{adolc}, and
Tapenade~\cite{tapenade}. Due to the language
features, the implementation of the AD packages is based on source
transformation for Fortran codes and operator overloading for C or C++ codes. 

When applying these AD-tools in a black box fashion to large-scale simulation codes,
one usually does not observe the theoretical runtime factors mentioned above, e.g., due to memory issues. 
Therefore, a structure exploiting approach, where AD is only applied to relevant parts of the code, as used in this work, is very often beneficial for the efficient calculation of the derivatives. This includes also the appropriate choice of the mode of AD. For the application considered here, where, e.g., derivatives of a matrix with respect to a parameter vector are required, the forward mode is the method of choice.

\begin{figure*}[]
\includegraphics[width=0.98\textwidth]{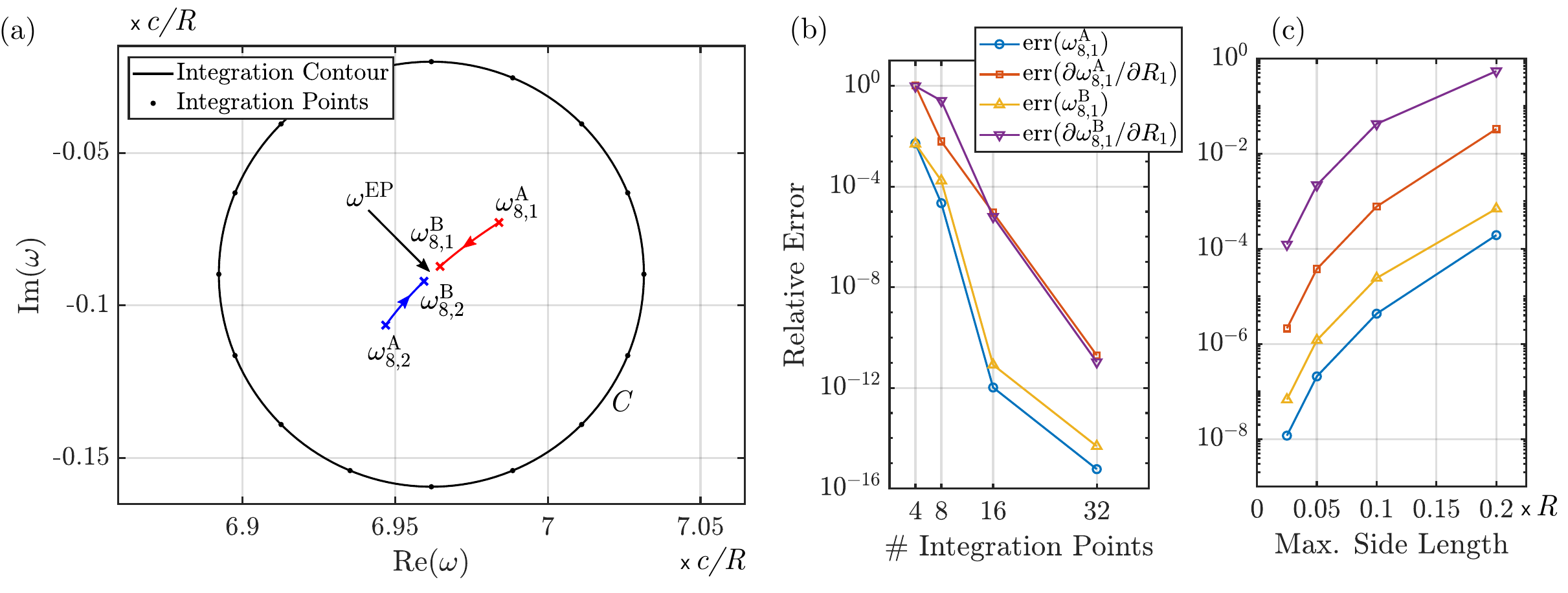}
\caption{\label{fig04}
Computing eigenfrequencies and their sensitivities near the EP.
(a)~Complex frequency plane with integration contour~$C$. The center of the contour is at ${\omega}^\mathrm{EP}$ and the radius is given by $0.01 \times \mathrm{Re}({\omega}^\mathrm{EP})$. Four eigenfrequencies from Fig.~\ref{fig03}\textcolor{blue}{(a)} are shown, where ${\omega}_{8,1}^\mathrm{A}$ and ${\omega}_{8,2}^\mathrm{A}$ or ${\omega}_{8,1}^\mathrm{B}$ and ${\omega}_{8,2}^\mathrm{B}$ are the result of a contour integration with a different $R_1$ in each case.
(b) Relative error $|{\omega}_{8,1}-{\omega}_{\mathrm{qex}}|/|{\omega}_{\mathrm{qex}}|$ with respect to the number of integration points on the contour $C$, where ${\omega}_{\mathrm{qex}}$ is the quasiexact solution computed with $64$ integration points. The maximum side length of the FEM mesh is $0.05 \times R$. 
(c)~Relative error with respect to maximum side length of the FEM mesh, where ${\omega}_{\mathrm{qex}}$ is the quasiexact solution computed with a maximum side length of $0.0125 \times R$. The number of integration points is $16$. Legend as before.
}
\end{figure*} 

\section{Application to a microdisk cavity}
We apply the presented framework to a two-dimensional optical microdisk cavity with two concentric layers of different refractive indices embedded in an open environment~\cite{Kullig_2018}. The system is sketched in Fig.~\ref{fig02}\textcolor{blue}{(a)}. The interest lies in the electric fields which are perpendicular to the cavity plane (TM polarization). This simplifies the time-harmonic Maxwell's equations~\eqref{eq:maxwell} to the scalar Helmholtz equation
\begin{align}
    \nabla^2 {E}_{m,l}^z(x,y) + n^2(x,y) \frac{{\omega}_{m,l}^2}{c^2}{E}_{m,l}^z(x,y) = 0, \label{eq:helmholtz}
\end{align}
where ${E}_{m,l}^z(x,y)$ is the $z$-component of the $l$-th eigenmode with the azimuthal mode number $m \in \mathbb{Z}$, the refractive index $n(x,y) = \sqrt{\epsilon_r \mu_r}$ describes the material in the region of interest, ${\omega}_{m,l} \in \mathbb{C}$ is the eigenfrequency, and $c$ is the speed of light.

\subsection{Semi-analytical solution}
Semi-analytical results for Maxwell's equations can be obtained in the case of two concentric dielectric disks \cite{Hentschel_2002}. As the system is rotational invariant, the eigenfrequencies are characterized by an azimuthal mode number $m\in\mathbb{Z}$. Therefore, combining the continuity conditions at the dielectric interfaces with the outgoing wave condition results in a conditional equation $S_m({\omega}_{m,l}) = 0$ for the eigenfrequency ${\omega}_{m,l}$. In the case of TM polarization, the function $S_m(\omega)$ is given by
\begin{widetext}
\begin{align}
\begin{split}
	S_m(\omega) =\; &n_2 J_m(k_1R_1)H^{(1)\prime}_m(k_3R)\left[H^{(2)\prime}_m(k_2R_1)H^{(1)}_m(k_2R)- H^{(1)\prime}_m(k_2R_1)H^{(2)}_m(k_2R)\right]\\ 
	 &- n_2^2 J_m(k_1R_1) H^{(1)}_m(k_3R)\left[H^{(2) \prime}_m(k_2R_1) H^{(1) \prime}_m(k_2R) - H^{(1)\prime}_m(k_2R_1)H^{(2)\prime}_m(k_2R)\right]\\ 
	 &- n_1 J^{\prime}_m(k_1R_1) H^{(1) \prime}_m(k_3R)\left[H^{(2)}_m(k_2R_1)H^{(1)}_m(k_2R) - H^{(1)}_m(k_2R_1)H^{(2)}_m(k_2R)\right]\\ 
	 &+ n_1n_2 J^{\prime}_m(k_1R_1)H^{(1)}_m(k_3R)\left[H^{(2)}_m(k_2R_1)H^{(1)\prime}_m(k_2R) - H^{(1)}_m(k_2R_1)H^{(2) \prime}_m(k_2R)\right],
 \end{split}\label{eq:SemiAnaMode}
\end{align}
\end{widetext}
where $k_j=n_j\omega/c$ and $J^{(1)}_m, J^{(2)}_m$, $H^{(1)}_m, H^{(2)}_m$ are Bessel-and Hankel functions of first and second kind and order~$m$, respectively.
The complex roots of $S_m(\omega)$ need to be determined numerically, e.g., by Newton's method, where an initial guess can be taken from a single disk with modified refractive index \cite{Hentschel_2002}.
With the implicit function theorem, the derivative of the eigenfrequency~${\omega}_{m,l}$ with respect to a parameter $p\in\lbrace n_1, n_2, n_3, R_1, R\rbrace$ can be calculated as
\begin{align}
	\frac{\partial {\omega}_{m,l}}{\partial p} = -\left[ \frac{\partial S_m}{\partial \omega} ({\omega}_{m,l})\right]^{-1}  \frac{\partial S_m}{\partial p}({\omega}_{m,l}). \nonumber
 \end{align}
The partial derivatives $\partial S_m/\partial \omega$ and $\partial S_m/\partial p$ are independent of the eigenfrequency. Therefore, they can be calculated analytically with a common computer algebra system from Eq.~\eqref{eq:SemiAnaMode}, see also Ref.~\cite{Binkowski_SourceCode_Sens_EPs}.

\begin{figure}[]
\includegraphics[width=0.49\textwidth]{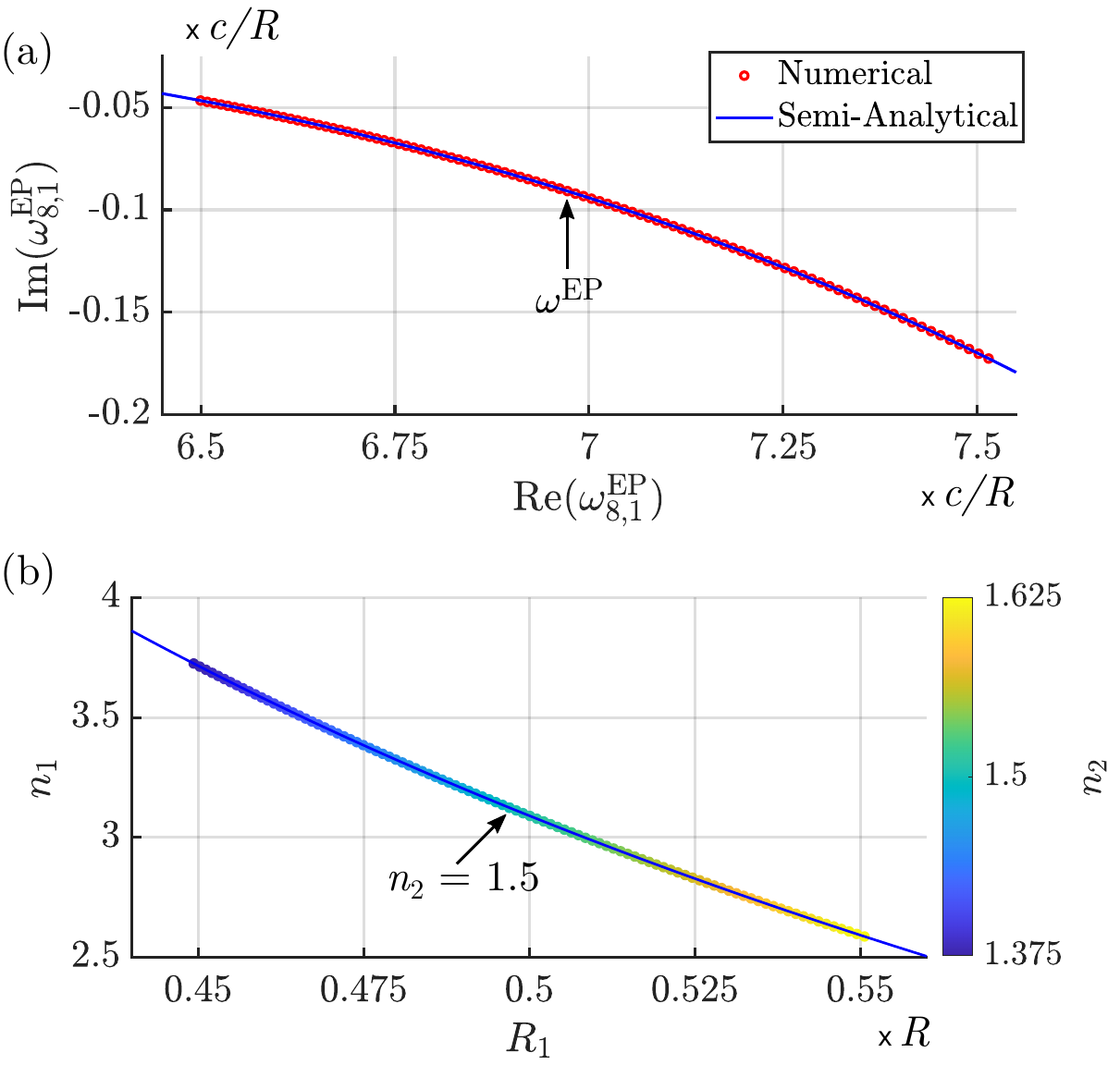}
\caption{\label{fig05}
Tracking the EP in the parameter space. 
(a) Numerical and semi-analytical results for the trajectory of the eigenfrequency ${\omega}^\mathrm{EP}$. The arrow indicates the eigenfrequency ${\omega}^\mathrm{EP}$ from Ref.~\cite{Kullig_2018}.
(b)~Corresponding parameter combinations. For each point in the parameter space, the parameter~$n_2$ is shifted by 0.0025.
}
\end{figure} 

\subsection{Numerical solution}
To numerically solve Eq.~\eqref{eq:helmholtz}, the software RPExpand~\cite{Betz_2021} is applied. The eigenfrequencies ${\omega}_{m,l}$ and the corresponding sensitivities $\partial {\omega}_{m,l}/ \partial p$ with respect to a parameter $p$ are computed by using the contour-integral-based framework. Here, the quantity $q(\omega)$ is the \mbox{$z$-component} of the electric field at the position $[x,y] = [0, 0.9 \times R]$ resulting from an $z$-polarized plane wave travelling in $y$-direction. RPExpand contains an interface to the software JCMsuite. The approach of AD is utilized within JCMsuite. 
Convergence with respect to the FEM parameters is ensured by refining the spatial mesh, where the degree of the polynomial ansatz functions is set to four. For computational efficiency, the mirror symmetries of the system are exploited. Further information on the numerical implementation can be found in Ref.~\cite{Binkowski_SourceCode_Sens_EPs}.

We study an EP of second order~\cite{Kullig_2018}. Two eigenmodes with an azimuthal mode number~$m=8$ and the corresponding eigenfrequencies coalesce at the parameter combination $n_1 = 3.1239791$, $n_2 = 1.5$, and $R_1 = 0.4970147 \times R$. The electric field intensity of a corresponding eigenmode ${E}_{8,1}^z$ is shown in Fig.~\ref{fig02}\textcolor{blue}{(b)}. The eigenfrequency is given by ${\omega}_{8,1}\approx{\omega}_{8,2} = (6.96185-0.089761i) \times c/R$. In the following, for a simpler notation, we denote this specific eigenfrequency by ${\omega}^\mathrm{EP}$.

The eigenfrequencies ${\omega}_{8,l}$ and their sensitivities $\partial {\omega}_{8,l} / \partial R_1$ near the EP are investigated. Figure~\ref{fig03}\textcolor{blue}{(a)} shows the trajectories of the real parts of the eigenfrequencies ${\omega}_{8,l}$ when the parameter $R_1$ is varied and $n_1$ and $n_2$ are fixed. For each radius $R_1$, a contour integral is computed leading to two eigenfrequencies. The superscript letters A and B describe two points in the parameter space that differ in the radius $R_1$. When $R_1$ is increased, i.e., going from ${\omega}_{8,1}^\mathrm{A}$ to ${\omega}_{8,1}^\mathrm{B}$ or ${\omega}_{8,2}^\mathrm{A}$ to ${\omega}_{8,2}^\mathrm{B}$, then the two eigenfrequencies converge. Figure~\ref{fig03}\textcolor{blue}{(b)} shows the imaginary parts and Fig.~\ref{fig03}\textcolor{blue}{(c)} and Fig.~\ref{fig03}\textcolor{blue}{(d)} show the corresponding sensitivities. 
The results can be compared with the semi-analytical solution ${\omega}_\mathrm{an}$. For ${\omega}_{8,1}^\mathrm{B}$, the relative error $|\mathrm{Re}({\omega}_{8,1}^\mathrm{B}-{\omega}_{\mathrm{an})}/\mathrm{Re}({\omega}_{\mathrm{an})}|$ is smaller than $7\times 10^{-7}$ and, for the corresponding imaginary part, the relative error is smaller than $5\times 10^{-5}$. The relative error for the underlying sensitivities is smaller than $2\times 10^{-3}$ for real and imaginary part. It can be observed that the sensitivities diverge in the vicinity of the EP. The divergence can be explained by the square root dependence of the eigenfrequencies on perturbations when the system is near the EP~\cite{Kato_1995}.

For all calculations corresponding to Fig.~\ref{fig03}, the contour $C$ shown in Fig.~\ref{fig04}\textcolor{blue}{(a)} with $16$ integration points and a maximum side length of the FEM mesh of $0.05 \times R$ are used. Figure~\ref{fig04}\textcolor{blue}{(a)} also sketches the locations of the eigenfrequencies ${\omega}_{8,1}^\mathrm{A}$, ${\omega}_{8,2}^\mathrm{A}$, ${\omega}_{8,1}^\mathrm{B}$, and ${\omega}_{8,2}^\mathrm{B}$ with the trajectories when $R_1$ is varied in the direction of the parameter for the EP. To demonstrate the accuracy of the approach, Fig.~\ref{fig04}\textcolor{blue}{(b)} shows the relative error of the absolute value of ${\omega}_{8,1}^\mathrm{A}$, $\partial {\omega}_{8,1}^\mathrm{A}/\partial R_1$, ${\omega}_{8,1}^\mathrm{B}$, and $\partial {\omega}_{8,1}^\mathrm{B}/\partial R_1$ with respect to the number of integration points on the contour $C$. Figure~\ref{fig04}\textcolor{blue}{(c)} shows the relative error with respect to the maximum side length of the FEM mesh. It can be observed that the relative error for the sensitivities is smaller for ${\omega}_{8,1}^\mathrm{A}$, which lies further away from the EP. The closer the two calculated eigenfrequencies lie to each other, the less accurate the results are.
With only $16$ integration points, it is possible to obtain a maximum relative error of smaller than $10^{-5}$ for the eigenfrequency sensitivities $\partial {\omega}_{8,1} / \partial R_1$ for both points A and B in the parameter space. Using a maximum side length of the FEM mesh of $0.05 \times R$ is sufficient to achieve a maximum relative error of smaller than $3\times10^{-3}$.

Note that, for simplicity, we only show the sensitivities $\partial {\omega}_{8,l} / \partial R_1$. The results for $\partial {\omega}_{8,l} / \partial n_1$ and $\partial {\omega}_{8,l} / \partial n_2$ exhibit a similar behavior.

\subsection{Exceptional surface}
We track the EP in the parameter space by using Newton's method with the sensitivities $\partial {\omega}_{8,l} / \partial n_1$ and $\partial {\omega}_{8,l} / \partial R_1$. We are looking for the zeros of the squared eigenfrequency splitting $({\omega}_{8,1}-{\omega}_{8,2})^2$. The reason for choosing the square of the splitting is the square root dependency of the eigenfrequencies near the EP. The starting point is the parameter combination corresponding to ${\omega}_{8,1}^\mathrm{B}$ and ${\omega}_{8,2}^\mathrm{B}$ shown in Fig.~\ref{fig03}\textcolor{blue}{(a)}, given by $n_1 = 3.1239791$, $n_2 = 1.5$, and $R_1 = 0.497004557\times R$. With a fixed $n_2=1.5$, a certain tolerance, and a few iterations of Newton's method, the two eigenfrequencies ${\omega}_{8,1}^\mathrm{EP} = (6.961993 - 0.089638i)\times c/R$ and ${\omega}_{8,2}^\mathrm{EP} = (6.961996 - 0.089642i)\times c/R$ with the parameters $n_1 = 3.123979246$ and  $R_1 = 0.497014753\times R$ are obtained. Then, the parameter $n_2$ is shifted by a constant value of $0.0025$ and Newton's method is restarted with the shifted $n_2$ together with the previously calculated values for the parameters $n_1$ and $R_1$, and with a new integration contour centered at the previously computed eigenfrequency. The results of this procedure are shown in Fig.~\ref{fig05}, where Fig.~\ref{fig05}\textcolor{blue}{(a)} contains the trajectory of ${\omega}^\mathrm{EP}$ and Fig.~\ref{fig05}\textcolor{blue}{(b)} shows the corresponding parameter combinations.

Although the sensitivities diverge near the EP, the numerical calculations of the sensitivities are still accurate enough for Newton's method to work very well. Therefore, the exceptional surface is identified as a curve in the parameter space spanned by $R_1$ and $n_1$. Figure~\ref{fig05} also contains the semi-analytical solution and it agrees with the numerical results.

\section{Conclusion}
Eigenfrequency sensitivities near EPs were investigated. 
For this purpose, we developed a framework based on contour integration with AD.
We applied a numerical implementation of the framework to an optical microdisk cavity with an EP of second order. It was shown that the eigenfrequency sensitivities near the EP can be captured accurately and efficiently. The sensitivities were applied to track the EP in the parameter space. A semi-analytical solution for the system was derived and used to validate the numerical results.

The presented contour-integral-based framework cannot only be applied to a benchmark problem as studied in this work, but also to any resonant system, e.g., with complex geometry or with an EP of arbitrary order. The combination of contour integration with AD makes the framework very general since the approach is essentially based on solving scattering problems. The treatment of the corresponding linear systems of equations is a standard task for state-of-the-art software packages used in the field of computational physics.

Computing eigenfrequency sensitivities near EPs can aid in optimizing setups for EP-based sensors. For example, exceptional surfaces \cite{Kim_OptExpr_2023} in higher dimensional parameter spaces can be identified, which can then be used to optimize an EP along such a surface to maximize sensitivity to a targeted perturbation while minimizing the response to other perturbations that may arise from fabrication tolerances or noise. For such an optimization scheme, the numerically accurate and efficient computation of eigenfrequency sensitivities with respect to specific system parameters is crucial.

\section*{Data and code availability}
Supplementary data tables and source code for the numerical experiments for this work can be found in the open access data publication~\cite{Binkowski_SourceCode_Sens_EPs}. Links to further AD tools and comprehensive information on AD can be found at www.autodiff.org. \vspace{0.8cm}

\section*{Acknowledgements}
We acknowledge funding
by the Deutsche Forschungsgemeinschaft (DFG, German Research Foundation) 
under Germany's Excellence Strategy - The Berlin Mathematics Research
Center MATH+ (EXC-2046/1, project ID: 390685689) and
by the German Federal Ministry of Education and Research
(BMBF Forschungscampus MODAL, project 05M20ZBM).

\end{document}